 \documentstyle[twocolumn,prl,floats,aps]{revtex}
\begin{document}
\draft

%
\twocolumn[\hsize\textwidth\columnwidth\hsize\csname @twocolumnfalse\endcsname

\title{ Two-Dimensional {\em t-J} Model at Low Electron Density}
\author{C. Stephen Hellberg\cite{steve} and Efstratios
Manousakis\cite{stratos}}
\address{
Department of Physics, Florida State University, Tallahassee, FL 32306-3016 }
\date{\today}
\maketitle
\begin{abstract}
\noindent
The phase diagram of the two-dimensional {\em t-J} model
is determined accurately
at low electron density
by a combination of analytic and numerical techniques.
The ground state exhibits three phases
in the limit of zero electron density:
an unpaired state at small $J/t$,
a gas of $s$-wave pairs for $2 < J/t \lesssim 3.4367$,
and a phase separated state at larger interaction strengths.
Bound states of larger clusters are never realized,
and the instabilities present at small densities are discussed.
\end{abstract}

%
]

Models of strongly-correlated electrons
have received renewed attention since the discovery of the
cuprate superconductors \cite{dagotto94}.
However a reliable method of calculating ground-state properties
of such models on
dense large-size
systems has not yet been demonstrated.

In this paper, we concentrate on the low {\em electron} density limit of the
two-dimensional
{\em t-J} model, one of the simplest models with strong correlation
\cite{manousakis91,emery90,lin91,putikka92}.
This limit appears far from the
small {\em hole} densities thought
to be applicable to cuprate superconductivity,
but it is important to calculate the entire phase diagram of the model.
A complete description of the low electron density limit,
where much can be calculated analytically,
yields a solid basis from which
to move into the potentially more interesting regions of the model.
Additionally, we introduce and test a numerical technique that we expect will
prove to be a powerful tool for investigating
strongly correlated models at higher electron densities.

The {\em t-J} Hamiltonian
\begin{equation}
	H
	= - t
		\sum_{\langle ij \rangle \sigma}
		( c_{i\sigma}^{\dagger}c_{j\sigma}
		+  c_{j\sigma}^{\dagger}c_{i\sigma} )
 +
		J
\sum_{\langle ij \rangle}
		( {\bf S}_{i} \! \cdot \! {\bf S}_{j} -
			\frac{n_i n_j}{4} )
\label{tj-ham}
\end{equation}
acts only on the constrained Hilbert space with no doubly occupied sites.
The operator
$c_{i\sigma}^{\dagger}$ creates an electron of spin $\sigma$ on site $i$,
${\bf S}_{i}$ is the spin operator, and
$n_i = \sum_\sigma c_{i\sigma}^{\dagger}c_{i\sigma}$.
The sum over $\langle ij \rangle$ enumerates nearest-neighbor bonds of a
square lattice.

As $J/t \rightarrow 0$,
the electrons separate to avoid interactions in the limit of zero density,
and the ground-state energy per particle is the kinetic
energy of free electrons, ${\cal E} _{N=1} = -4t$.
Assuming a macroscopic number of electrons,
the system phase separates completely in the large interaction strength limit,
and the 
energy per particle, ${\cal E} _{N=\infty} = -1.16934 J \pm 0.00003 J$,
comes purely from the interaction term in (\ref{tj-ham}).
This value is the result of the most accurate calculation of the
ground-state energy of the Heisenberg model \cite{runge92}
shifted by $-\frac{1}{4}J$ per bond.

For intermediate interaction strengths, the situation becomes more complicated.
It is well known that for $J/t > 2$, pairs of electrons bind into
an $s$-wave state
\cite{emery90,lin91}.
However, previous work only cursorily examined the transition
from this pairing phase
to the completely phase-separated state at large $J/t$.

This paper presents a thorough calculation of the 
phase diagram of the two-dimensional {\em t-J} model at zero
electron density.
We calculate the energy per particle of all potential
intermediate phases as a function of $J/t$.
We show that no intervening phase is stabilized between
the gas of $s$-wave pairs and the completely phase-separated state,
and we determine the phase boundaries accurately.
Finally we discuss possible instabilities at low densities.


The energy of a bound electron pair may be found analytically,
and we give the exact solutions below.
In the limit of large $J/t$, a pair binds into a singlet dimer
with an energy per particle of ${\cal E} _{N=2} = - \frac{1}{2} J$.
Since the chemical potential of a bound triplet is the same in
the strong interaction limit, we dismiss three-particle bound states
as a viable intermediate phase.

The next competitive intervening phase to consider is a gas of bound quartets.
Four electrons bind into a square configuration
with an energy 
${\cal E} _{N=4} = - \frac{3}{4} J$ per particle
in the large $J/t$ limit.
Since this value is significantly lower than the pair chemical potential,
quartets warrant careful scrutiny as a potential ground-state phase.
We use a Green's Function Monte Carlo
method to solve the four-body problem on finite lattices
and extrapolate to infinite system size.
Close to the critical $J/t$ for four electrons to bind,
the weakly bound quartets are very extended objects.
An accurate calculation of this critical interaction strength
on the infinite lattice
requires precise ground-state
energies of four electrons on very large lattices.
With the computational method described below,
we can calculate ground-state properties on lattices with dimensions
as large as
$20 \times 20$.
%
The results enable us to determine that
bound quartets are never energetically stabilized in a macroscopic system.



The paper is organized as follows:
The solutions of two and four particles on the infinite lattice are outlined.
We then calculate and discuss the low-density phase diagram.

The equations of motion of two electrons 
yield two bound states for sufficiently large
interaction strengths,
one with $s$-wave symmetry and one with $d_{x^2-y^2}$ 
symmetry \cite{lin91,mattis86,mattis94,boninsegni}.
The analysis is simplified by defining the reduced $s$-wave pair binding energy
$\delta _s = (-8t -E _s)/8 \geq 0$,
where $E _s$ is the total energy of the $s$-wave pair.
The reduced $d$-wave pair binding energy $\delta _d$ is defined similarly.

On the infinite lattice, 
the $s$-wave solution satisfies
\begin{equation}
\frac{1}{J} = \frac{1+\delta _s}{2}
\left[ 1 - \frac{\pi }{ 2} \left/
\mbox{\boldmath $K$}\left( \frac{1}{1+\delta _s} \right) \right. \right] ,
\label{js-exact}
\end{equation}
where
$\mbox{\boldmath $K$}(k)$ is the complete elliptic integral of the first kind
\cite{lin91}.
{}From now on, we express all energy scales in units of $t$.
This expression may be expanded close to the critical $J$ for pairs to bind.
To order $\delta _s$,
we find
\begin{equation}
\frac{1}{J}
	\approx \frac{1}{J^s _c} +\frac{ \pi}{2 }\frac{1}{ \log ( \delta _s / 8 ) } +
	\frac{1}{2}\delta _s +\frac{\pi}{4} \frac{\delta _s }{\log (\delta _s / 8)} +
	\cdots
\label{js-expand}
\end{equation}
where $J^s_c = 2$ is the critical interaction strength to bind $s$-wave pairs.
An expansion may also be made for large binding energy \cite{lin91,mattis94}.

After a similar analysis to that given in Ref.\ \cite{lin91},
the large-lattice limit of the $d$-wave solution reduces to
\begin{equation}
	\frac{1}{J}
= \frac{2 + \delta _d}{\pi} \mbox{\boldmath $E$}
\left( \frac{\sqrt{1+\delta _d}}{1 + \delta _d/2} \right)
- \frac{1 + \delta _d}{2} ,
\label{jd-exact}
\end{equation}
where $\mbox{\boldmath $E$}(k)$
is the complete elliptic integral of the second kind.
An expansion for small binding energy yields
\begin{equation}
\frac{1}{J} \approx
\frac{1}{J^d_c}
- \frac{\pi-2}{2\pi} \delta _d
-\frac{\delta_d^2 \log \delta_d}{4\pi} + \frac{6 \log 2 - 1}{8 \pi}  \delta_d^2
 + \cdots
\label{jd-expand}
\end{equation}
where the critical interaction strength for $d$-wave pairing is
$J^d_c = 2 \pi / (4 - \pi) \approx 7.32$.

We solve the four-electron problem with a version
of Green's Function Monte Carlo (GFMC)
\cite{boninsegni,hellberg,chen,liang90}.
We project a trial state onto the ground state by operating on it repeatedly
with the Hamiltonian.
This generates a series of increasingly accurate approximants to the
ground state labeled by integers $ |p \rangle = H^p | \Psi\rangle$.

It is useful to expand the trial state in terms of the exact eigenstates:
\begin{equation}
|\Psi\rangle =
a _0 | \Psi _0 \rangle +
a _1 | \Psi _1 \rangle +
a _2 | \Psi _2 \rangle +
\cdots
\label{expand}
\end{equation}
where $ | \Psi _0 \rangle$ is the ground state, $| \Psi _1 \rangle$ is the
first excited state, etc.,
and the $a _i$'s are the expansion coefficients.
Rewriting the projected states in this way, we see
\begin{equation}
   | p \rangle \propto | \Psi _0 \rangle
	+ \frac{a_1}{a_0} \left( \frac{ E_1 }{ E_0 } \right) ^p \! | \Psi_1 \rangle
	+ \frac{a_2}{a_0} \left( \frac{ E_2 }{ E_0 } \right) ^p \! | \Psi_2 \rangle
	+ \cdots
\end{equation}
where $E_i$ is the energy of the $i$'th eigenstate.
Thus $| p \rangle$ approaches the lowest eigenstate for large $p$ provided
$a_0 \ne 0$ and
$ | E_{i>0} | < |E_0 | $
for all excited state energies $E_{i>0}$, a condition
satisfied by the {\em t-J} model for $J > 0$.

Care is needed to choose a trial state having maximal overlap
with the true ground state.
We try to write a very general form for the trial state but restrict it to
be a spin singlet with zero total momentum.
We use the Jastrow-pairing state:
\begin{equation}
| \Psi \rangle =
\prod_{i<j}
f({\bf r}_{i } - {\bf r}_{j })
P_4 \prod_{\bf k}
(u_{\bf k} +
v_{\bf k} c^\dagger_{{\bf k} \uparrow} c^\dagger_{-{\bf k} \downarrow})
|0 \rangle
\label{trial_state}
\end{equation}
where $c^\dagger_{{\bf k} \sigma}$ is the 
Fermion creation operator and
$P_4$ projects the state onto the subspace with four particles
\cite{gros89}.
We use mixed boundary conditions
since the GFMC converges faster with a closed shell.

The Jastrow factor $f({\bf r})$
correlates {\em all} pairs of particles, yielding a
total-spin-singlet state.
We satisfy the {\em t-J} model's hard-core constraint by requiring
$f({\bf r}  =  {\bf 0}) = 0$.

The GFMC needs a positive-definite guiding function, which we take to be
\begin{equation}
\Psi^g =
\prod_{i<j}
f^g({\bf r}_{i } - {\bf r}_{j })
\prod_{i' \! , \, j'}
s^g({\bf r}_{i' \uparrow} - {\bf r}_{j' \downarrow}) .
\label{guide_state}
\end{equation}
In general $f^g({\bf r}) \not= f({\bf r})$.
Since it is not important to guide with a spin-singlet function, we use the
additional spin-dependent Jastrow factor $s^g({\bf r})$
in the guiding function.
The primed indices enumerate electrons of one spin.

Auspicious choices of trial and guiding states can drastically reduce the
statistical errors in the GFMC.
We use much more general and potentially far superior states than
previous work.
Since the electrons on a given size lattice only occupy a finite number of
locations {\bf r}, we let the factors in (\ref{trial_state}) and
(\ref{guide_state}) vary independently at each distance or wave vector
not related by symmetry.
We apply all rotation and mirror symmetries to the Jastrow factors,
but only the mirror symmetries about the axes and parity
to the Fermion pairing fields $u_{\bf k}$ and $v_{\bf k}$.
For example, the Fermion pairing fields may be any real linear combination
of an $s$ and $d_{x^2-y^2}$ pairing state,
but time-reversal breaking and $d_{xy}$ symmetries are excluded.

On a $20 \times 20$ lattice, each Jastrow factor has 400 parameters,
as do the Fermion pairing fields.
The symmetry restrictions reduce the 800 total parameters in each state to
172 independent parameters in the trial state
and 128 
in the guiding function.
To optimize the parameters, we minimize the variance of the local energy
\cite{umirigar88}.
For sufficiently large $J/t$, we find
the spin-independent Jastrow factors 
in (\ref{trial_state})
and (\ref{guide_state}) bind the electrons in the optimized state, while the
pairing fields $u_{\bf k}$ and $v_{\bf k}$ and the guiding function's
spin-dependent Jastrow factor
$s^g({\bf r})$ provide the internal correlation of the bound quartet.

\begin{figure}[bt]
\caption{
Plot of the GFMC output in the form of equation (\protect\ref{func1})
for four electrons on a $20 \times 20$ lattice with $J/t = 5.5$.
The data converge to the total ground-state energy at large power $p$.
The solid line is the fit to (\protect\ref{fit1}) taking five terms
in (\protect\ref{expand}).
The inset shows the scaling of
the total ground-state energy with linear system size $L$.
The fit to (\protect\ref{scale_eq}) is used to extrapolate to the
infinite system.
}
\label{power}
\end{figure}

The statistical errors increase exponentially with increasing power $p$,
so we use the GFMC output at small powers to extrapolate to
infinite power.
We assume that the expansion
(\ref{expand}) can be approximated by a small number
of terms, typically about five, and we fit the output
to determine the coefficients $a_i$ and energies $E_i$.
We include
enough terms in (\ref{expand}) to make the systematic error due to
omitting terms smaller than the statistical errors.

Previous work \cite{boninsegni,hellberg} fit the GFMC output
to
\begin{eqnarray}
\frac{\langle p | H | p \rangle}{\langle p | p \rangle}
& = &
\frac	{\langle \Psi | H^{2p+1} | \Psi \rangle}
			{\langle \Psi | H^{2p}   | \Psi \rangle}
\label{func1}
\\
& = & \sum_{i} |a_i|^2 E_i^{2p+1} \Bigl/ \sum_{i} |a_i|^2 E_i^{2p}
\; \; \stackrel{p \rightarrow \infty}{\longrightarrow} \; \; E_0 .
\label{fit1}
\end{eqnarray}
Fig.\ \ref{power} shows a nicely converged example in this form.
While this function has the advantage of converging to the ground-state energy
for large $p$, it is quite difficult to fit (\ref{fit1})
with more than two terms in (\ref{expand}).
We instead fit the simpler function
\begin{equation}
\frac{\langle\Psi|H^p|\Psi \rangle}{\langle \Psi | \Psi \rangle}
=
\sum_{i} |a_i|^2 E_i^p
\; \; \stackrel{p \rightarrow \infty}{\longrightarrow} \; \; |a_0|^2 E_0^p
\label{fit2}
\end{equation}
to determine all the coefficients $a_i$ and energies $E_i$.

For each value of $J/t$, we calculate the ground-state energy on square
lattices
of dimensions
$L \times L$, where $L$ ranges from $L=4$ to $L=20$.
We expect the energy of a bound state to converge exponentially with
linear system size.
To determine the infinite-lattice ground-state
energy $E (L = \infty)$, we fit the finite results to
\begin{equation}
E (L) = E (L = \infty) - D \exp(-L/\xi)
\label{scale_eq}
\end{equation}
where $E (L)$ is the ground-state energy on an $L \times L$ lattice,
and $D>0$ and $\xi>0$ are parameters.
The inset in Fig.\ \ref{power} shows
a sample
extrapolation with system size.

\begin{figure}[bt]
\caption{
Energy per particle for each phase 
at zero electron density.
The horizontal dotted line is the kinetic energy of free electrons, and the
dot-dashed line is the potential energy of the macroscopically
phase-separated system.
The short-dashed and long-dashed lines are the
energies of $s$-wave and $d$-wave
two-particle bound states, respectively.
Their intersections with the free electron energy are marked with diamonds,
and the intersection of the $s$-wave pair energy
with the phase-separated energy is marked with the square.
The circles with error bars are the four-particle energies,
and the solid line is the
best fit of the quartet binding energies to (\protect\ref{fit}).
The fit intersects the $s$-wave energy at the triangle.
}
\label{energy}
\end{figure}

The energies of each phase are shown in Fig.\ \ref{energy}.
The two-particle bound-state energies plotted are the exact expressions
obtained by solving the integral equations
(\ref{js-exact}) and (\ref{jd-exact}).
The $s$-wave pair binds with
a weak logarithm beyond the critical interaction strength
$J_c^s = 2$, while the $d$-wave binding energy turns on with a healthy linear
term for $J > J_c^d \approx 7.32$.
The four-particle energies are extrapolated to infinite system size.
Four-particle energies for $J \le 5.21$ are statistically indistinguishable
from the $s$-wave pair energies and are omitted.

Only three of the phases considered are realized as ground states.
For $J \le 2$, the electrons are free, minimizing their kinetic
energy.
At larger interaction strengths, they bind into $s$-wave pairs, and finally
the system undergoes macroscopic phase separation to form a Heisenberg
cluster.
The transition 
to complete phase separation occurs at $J_c^{ps} = 3.4367 \pm 0.0001$,
which is determined by equating the phase separated
energy with the solution of (\ref{js-exact}).
The error of the transition point is due to the uncertainty in the best
estimate of the ground state energy of the Heisenberg model
\cite{runge92}.
This value disagrees significantly
with previously published results \cite{emery90,lin91}.

It is obvious from Fig.\ \ref{energy} that four-particle bound states are
never stabilized in a macroscopic system.
To determine the critical interaction 
$J_c^q$ for two pairs of electrons to bind into a four particle state,
we fit the quartet binding energies close to the transition to the form
\begin{equation}
\frac{1}{J} = \frac{1}{J_c^q} + \frac{A}{\log \delta_{4} + B} + C \delta_{4}
\label{fit}
\end{equation}
where $A \geq 0$, $B$, and $C$ are parameters.
The binding energy is given by
$\delta_{4} = -(E _{N=4} - 2 E _s) \ge 0$, where $E _{N=4}$ and $E _s$ are
the total infinite-lattice ground-state energies of the four-particle
and $s$-wave pair states, respectively.
The form of the fitting function
is a generalization of the first three terms in the expansion
of the $s$-wave pair binding energy (\ref{js-expand}),
and includes the first two terms of the
$d$-wave pair binding energy (\ref{jd-expand}).

The transition for pairs to bind into quartets takes place at
$J_c^q = 5.07 \pm 0.15$.
The statistical error is large due to the singular nature of the fitting
function (\ref{fit}).
Still it is clear that the quartet phase is never energetically favorable
to the macroscopically phase separated state.
The fit to (\ref{fit}) is shown as the solid line in Fig.\ \ref{energy}.

Because the four-body energy per particle is 
higher than that of
the phase-separated state, we see no need to consider gases of
larger clusters as potential ground states.
In the limit of large particle number $N$, the energy per electron
of a finite bound droplet scales as
\begin{equation}
{\cal E} _N \approx {\cal E} _{N=\infty}
+ \alpha \bigl/ \sqrt{N} + \cdots ,
\label{cluster}
\end{equation}
where ${\cal E} _{N=\infty}$
is the bulk energy per particle of the
phase-separated state and $\alpha > 0$ is the surface
correction.
Since the surface correction is positive, the energies of large clusters
approach the phase-separated energy monotonically from {\em above}
with increasing droplet size,
and ${\cal E} _N >  {\cal E} _{N=\infty}$ for sufficiently large $N$.
Clearly this inequality is violated for $N=2$, but we have shown it
holds for $N=4$.
Assuming the energy per particle of clusters is a reasonably smooth
function of cluster size, the phase-separated state will have lower
energy per electron than all droplets with four or more particles.
Thus the zero-electron-density phase diagram consists only of the three
phases described above.

\begin{figure}[bt]
\caption{
Phase diagram for low electron density.
The phase-separation boundary is estimated from the two-body solution on
finite systems,
and the other lines
are calculated from expressions in Ref.\ \protect\cite{kagan94}.
All phase boundaries become less accurate with increasing electron density
per site $n$.
}
\label{phase}
\end{figure}

At low but non-zero electron densities,
both the phase-separated and $s$-wave pair states 
remain robust.
However, the phase of free electrons at $J \le 2$ 
is unstable to higher-angular-momentum pairings
similar to the Kohn-Luttinger effect in three dimensions
\cite{kagan94,chubukov93}.
Using expressions (20) and (21) in Ref.\ \cite{kagan94},
we find the phase diagram shown
in Fig.\ \ref{phase}, which differs substantially
from the phase diagram in that Reference.
As the electron density increases from zero,
the phase-separated and $s$-wave pairing states persist,
but the free-electron phase is unstable to $p$-wave pairing at small $J$ and
$d_{x^2-y^2}$ pairing at larger $J$.
The energy gaps in these higher-angular-momentum pairing states
increase from zero
with increasing density, but remain extremely small at the low densities
where the 
effect is valid.
For example, at electron density $n = 0.1$, all $p$ and $d$-wave energy
gaps are less than $10^{-7}$ in units of $t$.

In summary, we have calculated the complete phase diagram of the
two-dimensional
{\em t-J} model in the limit of low electron density.
The phase-separated state at large $J$ evaporates into a gas of $s$-wave pairs
as the interaction strength is reduced---Gases of larger finite clusters
are never stabilized.
The $s$-wave pairs unbind as $J$ is reduced further,
but the resulting Fermi liquid is unstable first to $d$-wave pairing and
finally to $p$-wave pairing.


We thank Y. C. Chen, S. W. Haas, A. V. Chubukov,
and E. Dagotto for useful conversations.
This work was supported by the Office of Naval Research Grant No.
N00014-93-1-0189.
The numerical
calculations were performed on the 256-node Paragon at Wright Patterson
Air Force Base, a Department of Defense (DoD) High Performance Computing (HPC)
Shared Resource Center, funded by the DoD HPC Modernization Plan.

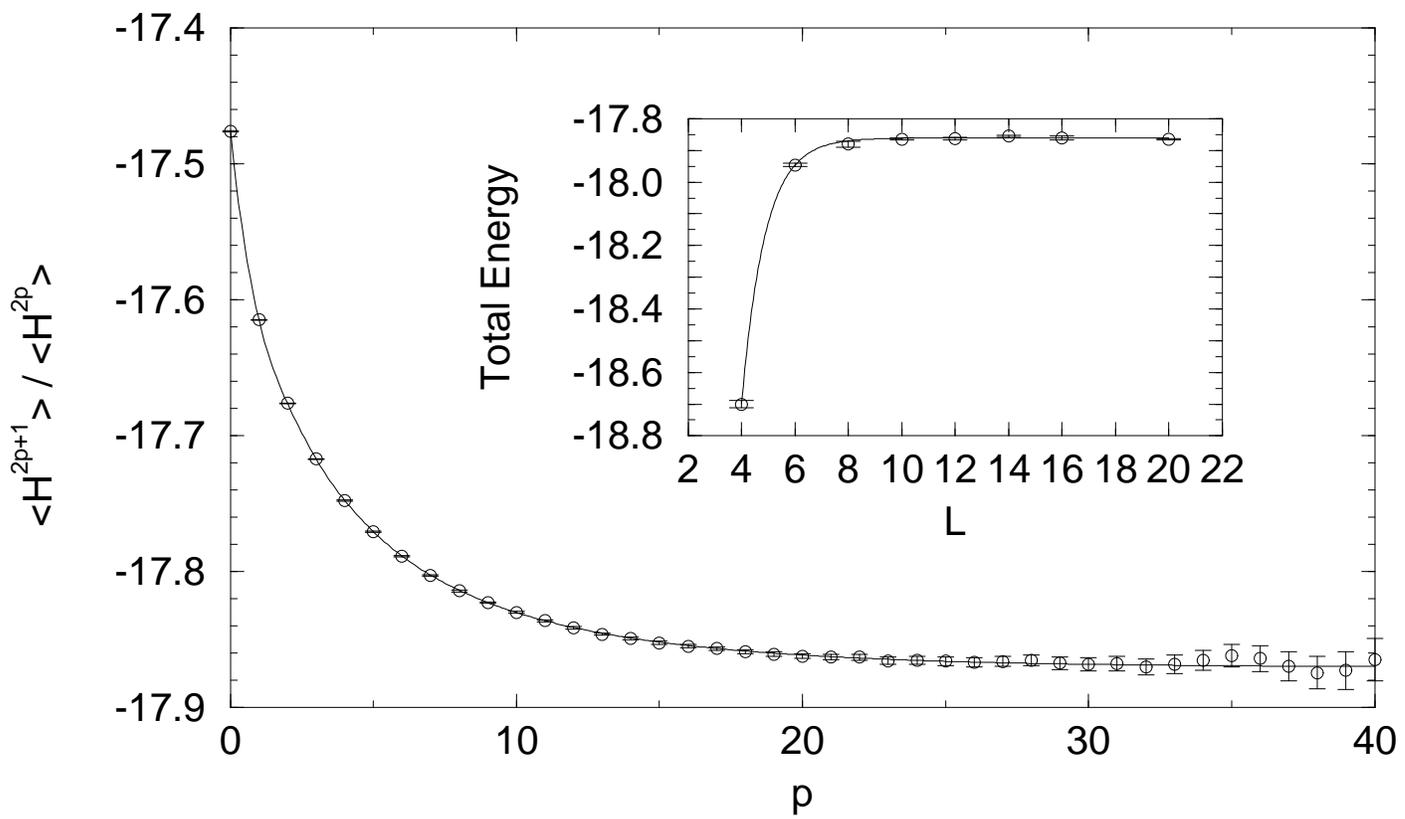
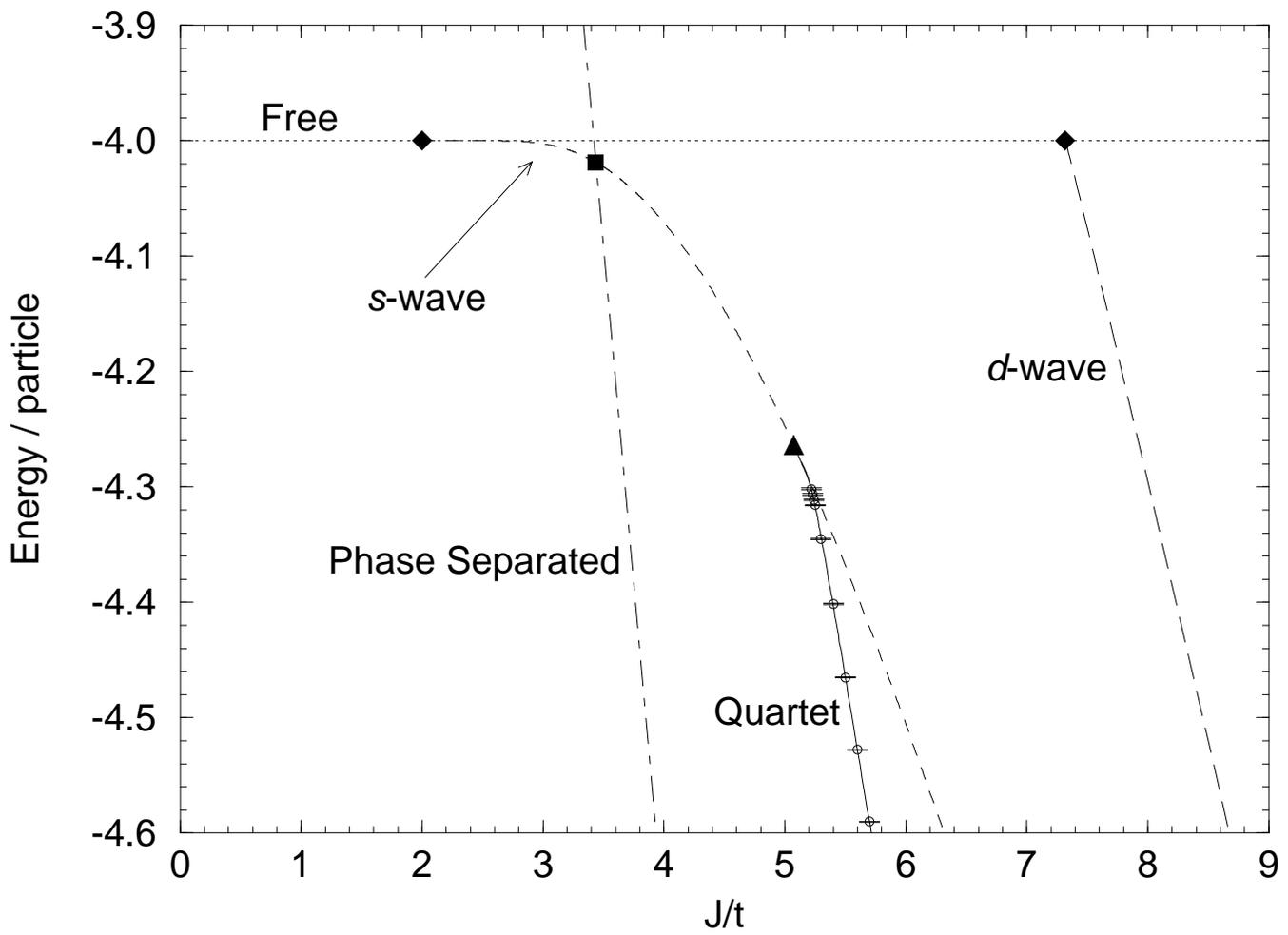
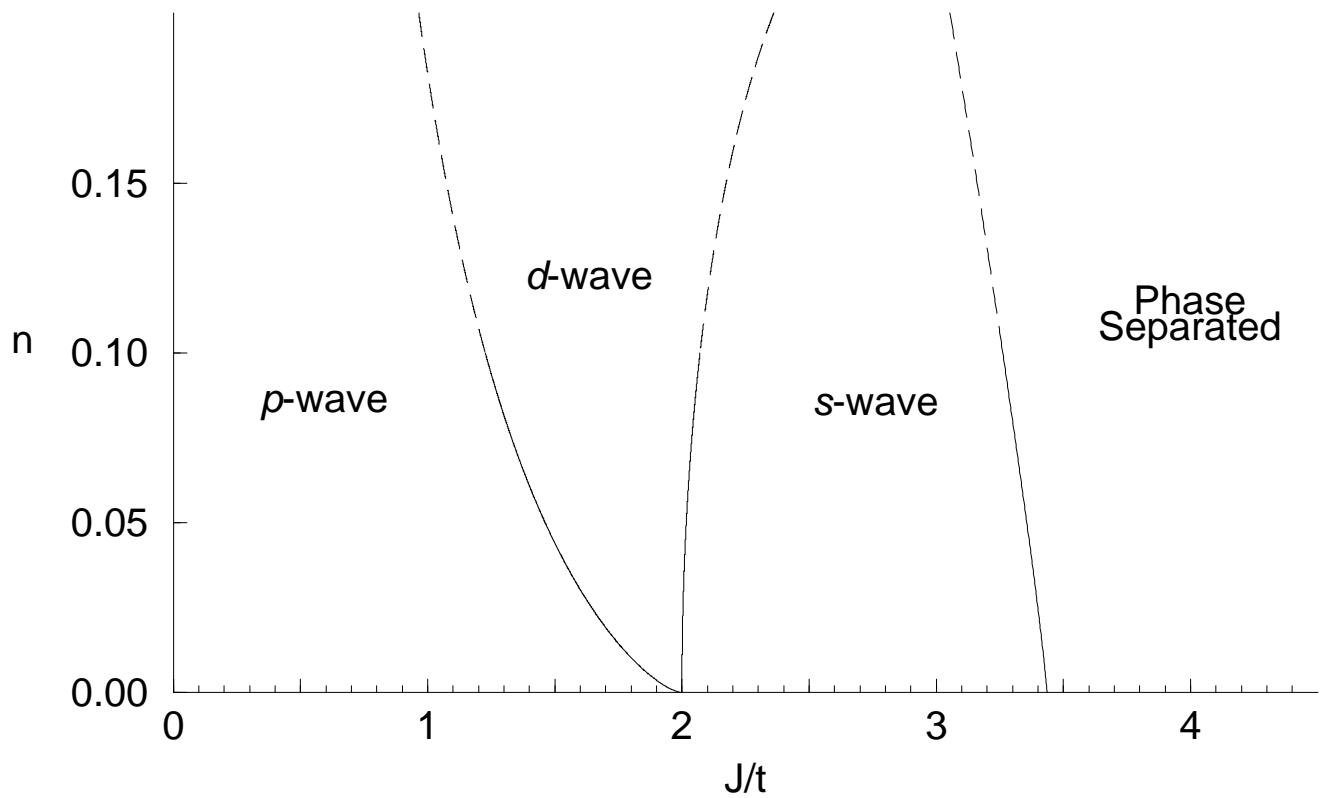
\end{document}